\documentclass{article}
\pdfoutput=1
\usepackage{amsmath,amssymb}
\usepackage{graphicx}
\usepackage{color}
\usepackage{mathtools}
\usepackage[table]{xcolor}
\usepackage{graphicx} 
\usepackage{subcaption}

\usepackage{array}

\newcommand{\ls}[1]
    {\dimen0=\fontdimen6\the\font\lineskip=#1\dimen0
     \advance\lineskip.5\fontdimen5\the\font
     \advance\lineskip-\dimen0
     \lineskiplimit=0.9\lineskip
     \baselineskip=\lineskip
     \advance\baselineskip\dimen0
     \normallineskip\lineskip\normallineskiplimit\lineskiplimit
     \normalbaselineskip\baselineskip
     \ignorespaces}

\begin{document}

\bibliographystyle{abbrv}

\title{Analysis of Boolean Functions based on Interaction Graphs and their influence in System Biology}
\author{Jayanta Kumar Das$^{1}$ ,Ranjeet Kumar Rout$^{1}$, Pabitra Pal Choudhury$^{1}$\\
\small $^{1}$ Applied Statistics Unit, Indian Statistical Institute, Kolkata-700108, India.\\
\small Email:dasjayantakumar89@gmail.com,
\small ranjeetkumarrout@gmail.com ,\\
\small pabitrapalchoudhury@gmail.com
}

\date{}
\maketitle

\thispagestyle{plain}
\setcounter{page}{1}
\ls{1.5}

\begin{abstract}
Interaction graphs provide an important qualitative modeling approach for System Biology. This paper presents a novel approach for construction of interaction graph with the help of Boolean function decomposition. Each decomposition part (Consisting of 2-bits) of the Boolean functions has some important significance. In the dynamics of a biological system, each variable or node is nothing but gene or protein. Their regulation has been explored in terms of interaction graphs which are generated by Boolean functions. In this paper, different classes of Boolean functions with regards to Interaction Graph with biologically significant properties have been  adumbrated.

\end{abstract}

\section{Introduction}
Biological components (such as genes, proteins etc.) are continuously interacting through paths and their interaction regulates  the system into complex global dynamic behavior $[1]$ and Biologists are currently wasting a lot of time and effort in searching for all of the available information form biological regulatory networks of biological components. Dynamics of the network can be described by recurrence of synchronous iteration of Boolean function which can be used to form Boolean Network. Again On the other hand topology of the network can be described by a sign directed graph. . An interaction graph talks about the positive and negative influences between components. A signed directed graph having one vertexes which considered to be components, indicates the static abstraction of Biological Network $[1, 2, 3]$.

Boolean functions have huge application in the theory of computer science, cellular Networks etc. $[4, 5, 6]$ and Boolean Networks in System Biology have been elaborately discussed in $[5]$. Boolean networks (BNs) are extensively used to model biological regulatory networks $[7, 8, 9, 10, 11]$ i.e. to study the interactions between Biological components such as genes, proteins etc. Each Boolean Network has some Biological Components which are independently represented by local logical Boolean functions and associates with a Boolean value for each component in Boolean Networks. All Boolean functions are not accurately reflecting the behaviors of Biological systems and it is imperative to recognize classes of Boolean functions with biologically relevant properties. A subset of Boolean functions having noble characteristics of dynamics of Boolean networks is constructed. These functions have significance for determining their potential in a model. One such notable class and their biological properties have been introduced by Kauffman$[7, 8]$ . To identify Boolean network, which are biologically relevant is a major problem as the number of Boolean functions and the size of the state space of Boolean networks are growing exponentially $[1]$ with the increase of components. Different technique such as classical analysis, model checking may be intractable with large complex systems. A number of operations can be carried out on Interaction graph to make biologically relevant predictions about a regulatory system and Interaction Graph can also be used for predicting qualitative aspects of system biology. Fundamental issues in the analysis of Interaction graphs are the enumeration of paths and cycles (feedback loops) and calculation of shortest positive or negative paths $[12, 13, 14]$.  Some static analysis of Boolean Networks through Interaction graph has been studied in $[1]$.

In this paper, analysis of the Boolean functions through interaction graph have been discussed by partitioning $n$-variable Boolean function into $2$ fixed bits. Here we present a slightly different approach from [1] with regards to the definition of Interaction graph.  Partitioning of a Boolean function into $2$ bits helps us to identify an edge or arc and cycle on interaction graph. Arcs and Cycles on Interaction Graph are basically responsible for static analysis of Boolean Network. First we will give a formal definition of Interaction Graph based on partitioning method and then classify the Boolean function based on Interaction Graph. In section $2$, decomposition technique is discussed and thereby Interaction Graph and their matrix representation are given. Section $3$, Boolean functions have been analyzed with regards to Interaction Graph and section $4$ deals with concluding remarks emphasizing the key factors of the entire analysis.

\section{Definition and Notations}
Given any Boolean function $f$( $ x_{1} $,$ x_{2} $,$ x_{3} $,....,$ x_{n} $ ) of $n$-variable is a mapping from $\{0,1\}^{n}\rightarrow\{0,1\} $ which are having the string of bit length $2^{n}$ bits. Decomposition of a Boolean function of $n$-variable is the segmentation of the function into $ 2^{n-1}$ functions with respect to inputs for all possible combination of fixed variables $n-1$. Output of each segmented function is two bits string which are fixed ($00,01,10,11$). Decomposition technique of any Boolean function with a single and $n-1$ variable has been given in section 2.1.

\subsection{$D_{i}(f)$ - Decomposition}
\noindent
$D_{i}$ -Decomposition of any Boolean function $f$ in input $x_{i}$ is the segmentation of $f$ into two functions $f_{0}^{i}$ and $ f_{1}^{i}$ which are defined by all possible inputs $x_i$ where $i\in\{1,2,3,...,n\}.$
  
 \begin{center}
   \resizebox{10cm}{!}
   {
    \centering
    $D_{i}(f) :\left \lbrace \begin{array}{c}
   
        f_{0}^i=x_1{},x_{2},x_{3},\ldots,x_{i-1},0,x_{i+1},\ldots, x_{n}\\
        f_{1}^i=x_1{},x_{2},x_{3},\ldots,x_{i-1},1,x_{i+1},\ldots, x_{n}
   				
   \end{array} \right\rbrace$
   }
 \end{center}
 
 The bit string representations of $f_{0}^i$ and $ f_{1}^i$ are called decomposition fragments of the $D_{i}(f)$ -Decomposition having the length of bits string for each decomposition fragment is $2^{n-1}$.
 To decompose a $n$-variable Boolean function from $i^{th}$ to $j^{th}$position having $(n-1)$ number of variables for each segment is defined as follow, 
  \begin{center}
     \resizebox{8cm}{!}
     {
      \centering
      $D_{i,\ldots,j}(f) :\left \lbrace \begin{array}{l}
   
          f_{0,\dots,0}^{i,\ldots j}=f(x_{1},\ldots,00\ldots00,\ldots ,x_{n})\\
                            \hspace{3cm} .\\
                            \hspace{3cm} .\\
                            \hspace{3cm}.\\
          f_{1,\ldots,1}^{i,\ldots j}=f(x_{1},\ldots,11\ldots11,\ldots,x_{n})
     \end{array} \right\rbrace$
     }
   \end{center}
 Where $i\:\: and\:\: j\in {1,2,3,...,n}$
 
 \textbf{Example 1.} Let consider a $3$-variable Boolean function $f_{21}(x_{1},x_{2},x_{3})$ with the bits string $00010101$. We have taken 2(n-1) variable at a time to decompose $f$ as it is 3-variable Boolean function. So there are $3$ decomposition fragments of the function  $f_{21}$ and they are shown below; 
 \begin{center}
   \resizebox{10cm}{!}
   {
   \centering
     $D_{23}(f) :\left \lbrace \begin{array}{c}
     
          f_{00}^{23}=11\\
          f_{01}^{23}=00\\
          f_{10}^{23}=10\\
          f_{11}^{23}=00\\
                           			
     \end{array} \right\rbrace$ $\:\:\:\:\:$
     $D_{13}(f) :\left \lbrace \begin{array}{c}
       
          f_{00}^{13}=11\\
          f_{01}^{13}=00\\
          f_{10}^{13}=10\\
          f_{11}^{13}=00\\
                             			
     \end{array} \right\rbrace$ $\:\:\:\:\:$
     $D_{12}(f) :\left \lbrace \begin{array}{c}
         
          f_{00}^{12}=10\\
          f_{01}^{12}=10\\
          f_{10}^{12}=10\\
          f_{11}^{12}=00\\
                               			
      \end{array} \right\rbrace$ 
 }
 \end{center}     
 Here $D_{23}(f)$ indicates decomposition of the function $f_{21}$ with regards to variable $x_{2}$ and $x_{3}$ and so on. The definition of Interaction Graphs with regards to decomposition technique and the analysis of Interaction Graph can be detected with the help decomposition fragments of any Boolean function.
   
 \subsection{Interaction Graph (I.G) of f}
 \noindent
 The Interaction graph of $f$, denoted by $G(f)=(V,E)$, is the sign directed graph on vertexes set $V\in \{1,2,\cdots n\}$ corresponds to nodes and edges set $ E \in \{+,-\}$, an arc (positive or negative) between nodes. For all $i, j \in V$, there exist an arc $i \longrightarrow j$ if and only if there exist at least one $D_{i,\ldots,j}(f)=01$ or $10$ in decomposition fragments for positive and negative arc respectively.
 
 \textbf{Example $2.$} Let consider three $3$-variable Boolean functions $f_{1} (x_{1},x_{2},x_{3})= x_{1} \bigvee ( x_{2} \bigwedge x_{3} ),\:\:  f_{2} (x_{1},\:\:x_{2},x_{3})=  x_{1} \bigwedge x_{2} \bigwedge x_{3} ,\:\:  f_{3} (x_{1},x_{2},x_{3})= ( x_{1} \bigwedge  x_{2} ) \bigwedge x_{3}$ with the bits string $00010101$, $00000001$ and $1001000 $ respectively. The Decomposition fragments of the three functions $f_{168},f_{128}$ and $f_{17}$ are shown below;
 
 \begin{center}
    \resizebox{12cm}{!}
    {
     $D_{23}(f_{1}/f_{2}/f_{3}) :\left \lbrace \begin{array}{c}
    
         f_{00}^{23}=00/00/11\\
         f_{01}^{23}=01/00/00\\
         f_{10}^{23}=00/00/00\\
         f_{11}^{23}=11/01/00\\
                          			
    \end{array} \right\rbrace$ $\:\:\:\:\:$
    $D_{13}(f_{1}/f_{2}/f_{3}) :\left \lbrace \begin{array}{c}
      
         f_{00}^{13}=00/00/10\\
         f_{01}^{13}=01/00/00\\
         f_{10}^{13}=00/00/10\\
         f_{11}^{13}=11/01/00\\
                            			
    \end{array} \right\rbrace$ 
    } \end{center}
    
     \begin{center}
            
    \resizebox{6cm}{!}
            {
        
        $D_{12}(f_{1}/f_{2}/f_{3}) :\left \lbrace \begin{array}{c}
            
             f_{00}^{12}=00/00/10\\
             f_{01}^{12}=00/01/00\\
             f_{10}^{12}=00/01/10\\
             f_{11}^{12}=01/01/00\\
                                  			
         \end{array} \right\rbrace$ }
      \end{center}
    
  Here three Boolean functions for $3$-variable and there are $3$ decomposition segments for each function. So there are total $3 \times 3=9$ decomposition segments. Output for each decomposition segments (first segment for function $f_{168}$, second segment for function $f_{128}$ and third segment for $f_{17}$ and so on are shown . To represent edges connectivity of these three functions (three functions represents corresponding three nodes $1$, $2$ and $3$ respectively) of the Interaction Graph of running Example $2$ is shown in Fig $1$.
 \begin{table}[ht]
        \centering
        \resizebox{7.5cm}{!}
        {
        \begin{tabular}{c c c}
        \includegraphics [scale=1]{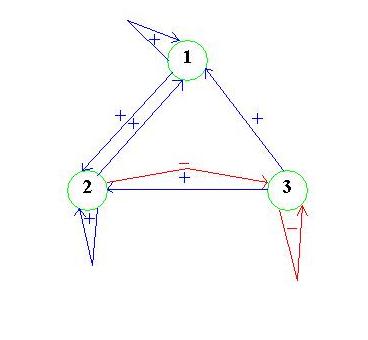} \\
        
         \end{tabular}
        }
        \begin{center}
        \textbf{Fig. 1.} I.G for functions $f_{168},f_{128}$ and $f_{17}$
        \end{center}
 \end{table}
   
\subsection{Matrix Representation of Interaction Graph}
\noindent
Since a graph is completely determined by specifying either its adjacency structure or its incidence structure, these specifications provide far more efficient ways of representing a large or complicated graph than a pictorial representation. As computers are more adept at manipulating numbers than at recognizing pictures, it is standard practice to communicate the specification of a graph to a computer in matrix form. We can represent node to node (vertex to vertex) connectivity of Interaction Graph by the matrices. For $n$ nodes size of the matrix will be $n \times n$ i.e. a square matrix $M=\left[a_{ij} \right]$  whose both the $n$ rows and $n$ columns correspond to the $n$ vertices shown in TABLE $1$ such that \\
\begin{center}
$a_{ij}= \left \lbrace \begin{array}{c}
1, $ if ith node is connected to jth node by positive edge $\\
-1, $ if ith node is connected to jth node by negative edge $\\
0, $ otherwise $\:\:\:\:\: \:\:\:\:\: \:\:\:\:\:\:\:\:\:\: \:\:\:\:\: \:\:\:\:\:\:\:\:\: \:\:\:\:\: \:\:\:\:\:\:\:\:\:\: \:\:\:\: \\
\end{array} \right\rbrace$ 
\end{center}
As Interaction Graph is signed directed Graph and direction of edges will be ith row to jth column ($i\longrightarrow j$) as each column is considered an individual Boolean function from node $1$ to node $n$, then each row from $1$ to $n$ represents the output (1 for 01, -1 for 10 and 0 for 11 or 00) of decomposition segment 1, segment $2$ \ldots segment $n$ respectively and vice versa . So the value of each cell will be $1,$ or $-1,$ or $0$.

\begin{table}[th]
\centering
\caption{Representation of $n \times n$ Matrix}
{\begin{tabular}{|c|c|c|c|c|c|c|c|c|c|c|}
\hline
\textbf{$i\downarrow j\longrightarrow $} & \textbf{1} & \textbf{2} & \textbf{3} & \textbf{4} & . & . & .& \textbf{n-2} & \textbf{n-1} &\textbf{n}\\
\hline 
\textbf{1} &  &   &  & & & & &  & &  \\
\hline 
\textbf{2} &  &   &  &  & & & &  & & \\
\hline 
\textbf{3} &  &   &  & & & & &  & &  \\
\hline 
\textbf{4} &  &   &  & & & & &  & &  \\
\hline 
.&  &   &  & & & & &  & &  \\
\hline
.&  &   &  & & & & &  & &  \\
\hline
.&  &   &  & & & & &  & &  \\
\hline
\textbf{n-2} &  &   &  &  & & & &  & &  \\
\hline
\textbf{n-1} &  &   &  &  & & & &  & &  \\
\hline
\textbf{n} &  &   &  &  & & & &  & &  \\
\hline
\end{tabular} }
\end{table} 
We represent two separate Matrixes i.e. Positive Matrix $(M+)$ and Negative Matrix $(M-)$ for positive edges and negative edges connectivity among nodes respectively for running Example $2$ shown below in Table $2$.
\begin{table}[th]
\centering
 \caption{Representation of $3 \times 3$ both Positive Matrix (M+) and Negative Matrix (M-)}
 {
  \begin{tabular}{|c|c|c|c|c|c|c|c|}
  \hline
  \multicolumn{4}{|c}{M+} & \multicolumn{4}{|c|}{M-} \\
  \hline
   $i\downarrow j\longrightarrow $ & \textbf{1} & \textbf{2} & \textbf{3} & $i\downarrow j\longrightarrow $ & \textbf{1} & \textbf{2} & \textbf{3} \\ \hline
   \textbf{1} & 1 & 1 & 0 & \textbf{1} & 0 & 0 & 0\\ \hline
   \textbf{2} & 1 & 1 & 0 & \textbf{2} & 0 & 0 & -1\\ \hline
   \textbf{3} & 1 & 1 & 0 & \textbf{3} & 0 & 0 & -1\\ \hline
  \end{tabular} }
  \end{table} 
 From $M+$, we can see that there exist  paths from node $1$ to node $1$ (self-loop), node $1$ to node $2$ and node $1$ to node $3$ which are represented in column $1$ and so on. And from $M-$ there exist paths node $2$ to node $3$ and node $3$ to node $3$.

\section{Analysis of Boolean Function}
This section provides analysis of Boolean functions with regards to their symmetrical Interaction graphs. For each case we classify the two sets of Boolean functions i.e. Positive Boolean Functions (PBF) and Negative Boolean Function (NBF) both binary and decimal value (DV) having similar Interaction graphs separately with positive edges and negative edges respectively.

\subsection{Only Positive or Negative Edges/Cycles in I.G}
\noindent
The Interaction graphs $G(f)$ have either only positive edges and positive cycles if $D_{i, \ldots ,j}(f)= 01$ or only negative edge and negative cycle if $D_{i, \ldots ,j}(f)= 10$ for all $(i, j)\in{1,2,3,..,n}$. Thus the Graph $G(f)$ using this type of functions may not always have a path $i\longrightarrow j\in G(f)$ and thereby may not always cycles of any length. List of functions (for n=2, 3 and 4 variable) which are satisfied this condition are shown in Table $3$ separately for positive and negative functions.
For $n=2$ there are total $4+4=8$ functions, for $n=3$ there are total $18+18=36$ functions and for $n=4$ there are total $166+166=332$ functions.

\textbf{Example 3:} Fig. 2(a) shown Interaction Graph of $3$ Boolean functions $f_{128},f_{168}$ and $f_{192}$ having positive edges only. 
  
\textbf{Example 4:} Fig. 2(b) shown Interaction Graph of $3$ Boolean functions $f_{23},f_{51} $ and $f_{3}$ having negative edges only.
  \begin{table}[ht]
     \centering
     \resizebox{10cm}{!}
     {
     \begin{tabular}{c c c}
     \includegraphics [scale=1]{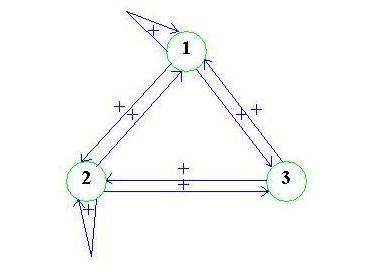} & \includegraphics [scale=1]{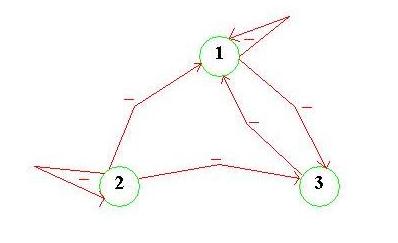}\\
     
     {\fontsize{1cm}{1cm}\selectfont (a)}&
     {\fontsize{1cm}{1cm}\selectfont (b)}\\
      \end{tabular}
     }
     \begin{center}
     \textbf{Fig. 2.}(a) I.G for functions $f_{128},f_{168} $ and $f_{192}$, (b) I.G for functions $f_{23},f_{51}$ and $f_{3}$.
     \end{center}
 \end{table}
 
 \subsection{All Positive or All Negative Edges/Cycles in I.G (complete graph)}
  \noindent
  The Interaction graphs $G(f)$ (Complete I.G) have either all positive edge and positive cycle iff $D_{i, \ldots ,j}(f)= 01$ or all negative edge and negative cycle iff $D_{i, \ldots ,j}(f)= 10$ for all $(i, j)\in{1,2,3, \cdots ,n}$. Thus the Graph $G(f)$ using this type of functions always have a path $i\longrightarrow j\in G(f)$ and thereby cycles of any length. List of functions (for n=2, 3 and 4 variable) which are satisfied this condition are shown in Table $4$ separately for positive and negative functions. For $n=2$ there are total $2+2=4$ functions, for $n=3$ there are total $9+9=18$ functions and for $n=4$ there are total $114+114=228$ functions.

  \textbf{Example 5:} Fig. 3(a) shown Interaction Graph of $3$ Boolean functions $f_{128},f_{168}$ and $f_{200}$ having positive edges only. 
  
  \textbf{Example 6:} Fig. 3(b) shown Interaction Graph of $3$ Boolean functions $f_{55},f_{21}$ and $f_{7}$ having negative edges only.
  \begin{table}[ht]
       \centering
       \resizebox{10cm}{!}
       {
       \begin{tabular}{c c c}
       \includegraphics [scale=1]{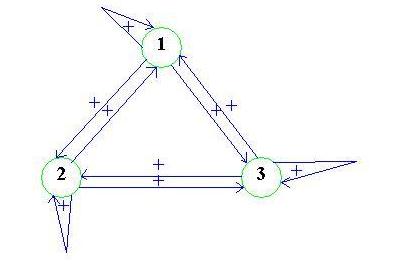} & \includegraphics [scale=1]{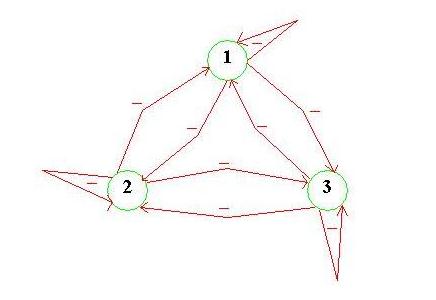}\\
       
       {\fontsize{1cm}{1cm}\selectfont (a)}&
       {\fontsize{1cm}{1cm}\selectfont (b)}\\
        \end{tabular}
       }
       \begin{center}
       \textbf{Fig. 3.}(a) I.G for functions $f_{128},f_{168}$ and $f_{200}$, (b) I.G for functions $f_{55},f_{21} $ and $f_{7}$.
       \end{center}
  \end{table}
  
\subsection{Nested Canalizing Functions (NCFs) with I.G}
\noindent
Not all Boolean functions reflect the behavior of biological systems and it is imperative to recognize the biologically relevant Boolean functions. One such class of Boolean functions is nested canalyzing function having small limit cycles and small average height in state space graph. In order to reduce the chaotic behavior and to attain stability in the gene regulatory network, nested Canalizing Functions (NCFs) are best suited. NCFs and its variants have a wide range of applications in systems biology $[15, 16, 17, 18]$.  So identification of all $n$-variable NCFs will be helpful for studying Boolean networks and hence biological networks.

If the Interaction graph $G(f)$ has no cycle, then iteraction graph $[1]$ has a unique fixed point. Nested canalizing functions carry special characteristics of an Interaction Graph. NCFs are connected to all components with self-loop in I.G.  That’s why all the nested canalizing Boolean functions can be used to generate graph with cycle having both positive and negative edges simultaneously. Nested Canalizing functions $[18]$ which are satisfied these conditions are shown in Table $5$. For $n=2$-variable there are total $8$ functions, for $3$-variable there are total $64$ functions.

 \textbf{Example 7:} Fig. 4. shown Interaction Graph of $3$ Nested Canalizing functions $f_{1},f_{8}$ and $f_{47}$ having three positive edges and six negative edges.\\
\begin{table}[ht]
       \centering
       \resizebox{7.5cm}{!}
       {
       \begin{tabular}{c c c}
       \includegraphics [scale=1]{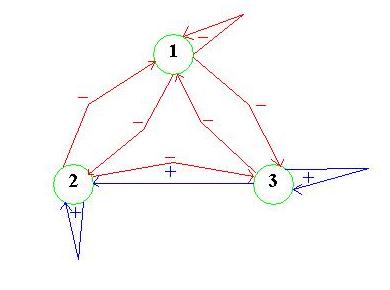} \\
       
        \end{tabular}
       }
       \begin{center}
       \textbf{Fig. 4.} I.G for functions $f_{1},f_{8}$ and $f_{47}$
       \end{center}
  \end{table}
  
 \section{Conclusion}
 In this paper, an attempt has been made for designing interaction graphs using Boolean function decomposition and various classes of Boolean functions are obtained to model a biological system with the help of interaction graph. In this method, parallel edges are not counted between two consecutive nodes for an Interaction Graph. Further study can be extended for counting the number of Boolean functions for $n-$variable and their applications towards static analysis of biologically regulated network. By knowing the functions, which are used to represent the genes/proteins, we can predict the characteristics of these functions and thereby help to the understanding of different biological networks through the pathways.

\begin{table}[th]
\centering 
\caption{Functions List For Section 3.1}
{
\centering
       \resizebox{6cm}{!}
       {
\begin{tabular}{|c|c|c|c|c|}
\hline
\textbf{VARIABLE} & \textbf{PBF} & \textbf{DV} & \textbf{NBF} & \textbf{DV}\\
\hline 
& 1000 & 8  & 0111 & 7  \\
n=2 & 1010 & 10 & 0101 & 5 \\
& 1100 & 12 & 0011 & 3 \\
& 1110 & 14 & 0001 & 1 \\
\hline 
& 10000000 & 128 & 01111111 & 127 \\
& 10001000 & 136 & 01110111 & 119 \\
& 10100000 & 160 & 01011111 & 95  \\
& 10101000 & 168 & 01010111 & 87 \\ 
& 10101010 & 170 & 01010101 & 85 \\  
& 11000000 & 192 & 00111111 & 63 \\
& 11001000 & 200 & 00110111 & 55 \\
n=3 & 11001100 & 204 & 00110011 & 51 \\
& 11100000 & 224 & 00011111 & 31 \\
& 11101000 & 232 & 00010111 & 23 \\
& 11101010 & 234 & 00010101 & 21 \\ 
& 11101100 & 236 & 00010011 & 19 \\
& 11101110 & 238 & 00010001 & 17 \\ 
& 11110000 & 240 & 00001111 & 15 \\
& 11111000 & 248 & 00000111 & 7 \\
& 11111010 & 250 & 00000101 & 5 \\
& 11111100 & 252 & 00000011 & 3 \\
& 11111110 & 254 & 00000001 & 1 \\
\hline
& 1000000000000000 & 32768 & 0111111111111111 & 32767 \\
& 1000000010000000 & 32896 & 0111111101111111 & 32639 \\
& 1000100000000000 & 34816 & 0111011111111111 & 30719 \\
&.&.&.&. \\
n=4 &.&.&.&. \\
&.&.&.&. \\
& 1110111011001000 & 61128 & 0001000100110111 & 4407 \\
& 1110111011001100 & 61132 & 0001000100110011 & 4403 \\
& 1110111011100000 & 61152 & 0001000100011111 & 4383 \\

\hline
\end{tabular}
}}
\end{table}

\begin{table}[th]
\centering
\caption{Functions List For Section 3.2}
{
       \resizebox{5cm}{!}
       {
\begin{tabular}{|c|c|c|c|c|}
\hline
\textbf{VARIABLE} & \textbf{PBF} & \textbf{DV} & \textbf{NBF} & \textbf{DV}\\
\hline 
n=2 & 1000 & 8 & 0111 & 7 \\
 & 1110 & 14 & 0001 & 1 \\
\hline
& 10000000 & 128 & 01111111 & 127 \\
& 10101000 & 168 & 01010111 & 87 \\
& 11001000 & 200 & 00110111 & 55 \\
& 11100000 & 224 & 00011111 & 31 \\
& 11101000 & 232 & 00010111 & 23 \\
n=3 & 11101010 & 234 & 00010101 & 21 \\
& 11101100 & 236 & 00010011 & 19 \\
& 11111000 & 248 & 00000111 & 7 \\
& 11111110 & 254 & 00000001 & 1 \\
\hline
& 1000000000000000 & 32768 & 0111111111111111 & 32767 \\
& 1000100010000000 & 34944 & 0111011101111111 & 30591 \\
& 1010000010000000 & 41088 & 0101111101111111 & 24447 \\
&.&.&.&. \\ 
n=4 &.&.&.&. \\
&.&.&.&. \\
& 1111100011100000 & 63712 & 0000011100011111 & 1823 \\
& 1111100011101000 & 63720 & 0000011100010111 & 1815 \\
& 1111100011110000 & 63728 & 0000011100001111 & 1807 \\
\hline
\end{tabular}}
}
\end{table}

\begin{table}[th]
\centering
\caption{Functions List For Section 3.3}
{
       \resizebox{9cm}{!}
       {
\begin{tabular}{|c|c|c|c|c|c|c|c|c|}
\hline
\textbf{VARIABLE} & \textbf{DV} & \textbf{BF} & \textbf{DV} & \textbf{BF} & \textbf{DV} & \textbf{BF} & \textbf{DV} & \textbf{BF}\\
\hline 
& 1 & 0001 & 8 & 1000 
& 2 & 0010 & 11 & 1011 \\
n=2 & 4 & 0100 & 13 & 1101 
& 7 & 0111 & 14 & 1110 \\
\hline
& 1 & 00000001 & 2 & 00000010 
& 4 & 00000100 & 7 & 00000111 \\
& 8 & 00001000 & 11 & 00001011 
& 13 & 00001101 & 14 & 00001110 \\
& 16 & 00010000 & 19 & 00010011 
& 21 & 00010101 & 31 & 00011111 \\
& 32 & 00100000 & 35 & 00100011 
& 42 & 00101010 & 47 & 00101111 \\
& 49 & 00110001 & 50 & 00110010 
& 55 & 00110111 & 59 & 00111011 \\
& 64 & 01000000 & 69 & 01000101  
& 76 & 01001100 & 79 & 01001111 \\
& 81 & 01010001 & 84 & 01010100 
& 87 & 01010111 & 93 & 01011101 \\
n=3 & 112 & 01110000 & 115 & 01110011 
& 117 & 01110101 & 127 & 01111111 \\
& 128 & 10000000 & 138 & 10001010 
& 140 & 10001100 & 143 & 10001111 \\
& 162 & 10100010 & 168 & 10101000 
& 171 & 10101011 & 174 & 10101110 \\
& 176 & 10110000 & 179 & 10110011 
& 186 & 10111010 & 191 & 10111111 \\
& 196 & 11000100 & 200 & 11001000 
& 208 & 11010000 & 213 & 11010101 \\
& 220 & 11011100 & 223 & 11011111 
& 224 & 11100000 & 234 & 11101010 \\
& 236 & 11101100 & 239 & 11101111 
& 241 & 11110001 & 242 & 11110010 \\
& 244 & 11110100 & 247 & 11110111 
& 248 & 11111000 & 251 & 11111011 \\
& 253 & 11111101 & 254 & 11111110 
& 205 & 11001101 & 206 & 11001110 \\
\hline
\end{tabular}}
}
\end{table}

\end{document}